# INFLUENCE OF NITROGEN IMPURITIES ON AN ELECTRON-EXCITED HELIUM ATOM CONCENTRATION IN THE SELF-SUSTAINED NORMAL DC GLOW DISCHARGE AT ATMOSPHERIC PRESSURE

V.I. Arkhipenko, A.A. Kirillov, L.V. Simonchik, S.M. Zgirouski

*Institute of Molecular and Atomic Physics NASB, Minsk, Belarus; simon@imaph.bas-net.by*

The influence of nitrogen impurities on the parameters of the self-sustained normal dc glow discharge at atmospheric pressure was studied. The concentrations of the low-excited helium atoms in states $2^1s$, $2^1p$, $2^3s$ and $2^3p$ were determined in the atmospheric pressure glow discharge in helium (99.98%He) and in helium with a nitrogen admixture. It was shown that the adding of nitrogen into helium (less 5%) leads to the increase of both interelectrode gap voltage and gas temperature. At the same time the drastically reduction of concentration of the low-excited helium atoms in cathode region even at nitrogen admixture of 0.5 % is occurred.

## I. INTRODUCTION

There is large increasing interest in atmospheric pressure glow discharges (APGD) because they can be used for a wide range of technological applications without the need of vacuum systems. As a rule, helium in mixture with other gases is used as working gas in discharges at atmospheric pressure [1, 2]. However, at present there isn't a complete understanding of the gas discharge physics in gas mixtures at atmospheric pressure. As it was shown in [2], for example, a small addition of other gases (~1%) into working gas of barrier discharge plays a significant role in stability of this discharge. On the base of numerical calculation it is approved in [3] that even the residual gas admixtures with concentration $\sim 0.5 \cdot 10^{-4}$ % exert a significant influence on both the plasma composition and the gas discharge parameters in atmospheric pressure discharge.

The self-sustained normal dc APGD in helium [4] is a convenient object for investigations of the kinetics of the weakly ionized nonequilibrium plasma with complicated composition. The low-excited levels of helium are participated at many plasma chemical reactions. Therefore their concentrations are one of major parameters of nonequilibrium plasma of the APGD. In present work the concentrations of the low-excited helium atoms in states $2^1S$, $2^1P$, $2^3S$ and $2^3P$ in the APGD in helium (99.98%He) and in helium with a nitrogen admixture (He 99.5% + $N_2$ 0.5%) were determined.

## II. EXPERIMENTAL SETUP

The self-sustained normal dc APGD was created in the pressurized chamber between two electrodes: the weakly rounded tungsten anode (diameter 6 mm) and flat circular cooper cathode (diameter 30 mm). The intensive water cooling of cathode was ensured due to its special design.

The interelectrode gap was about 4 mm. A weak flow of helium through the discharge chamber (less 1 litre/min) was provided. The impurity concentrations in helium flow ($H_2$, $N_2$, $O_2$, Ar, $CO_2$, CO, Ne, $H_2O$) did not exceed 0.02 %. The mixtures of helium with nitrogen were prepared in gas-cylinder in advance.

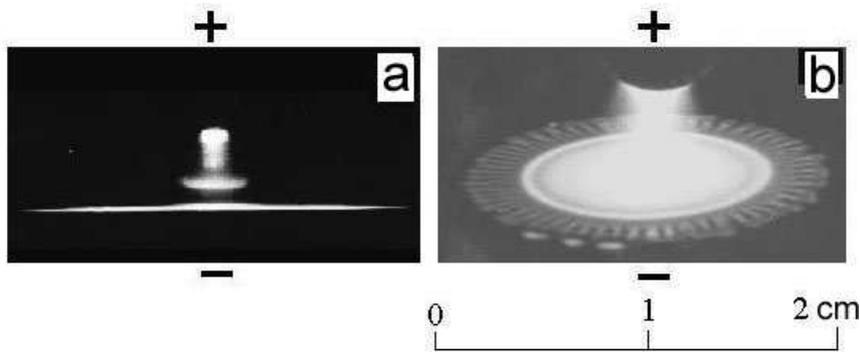

**Fig. 1.** *Photographs of the APGD in helium in the planes (a) parallel and (b) tilted with respect to the cathode surface.*

As it is seen in Fig. 1a, the structure of the glow discharge is as follows: a thin (less than 1-mm thick) disc of glow emission resides above the cathode surface, whereas the glowing column is adjacent to anode. There is the Faraday dark space between these glow regions. The anode surface is covered with a glowing layer. The positive column contracts to a diameter of about 3–5 mm.

The voltage- current characteristic of the APGD (Fig. 2) in case of a water-cooled cathode is slowly rising in the current and voltage ranges 0.05–15 A and 180–250 V respectively [5]. More quick growth of the interelectrode voltage, when discharge current is increasing from 1 A to 15 A, is the result of the insufficiency of cathode cooling. As was shown in [6], the potential profile along the discharge gap has a shape typical for classical glow discharges [7]: the major voltage drop (150–200 V) occurs in the cathode region, whereas the minor voltage drop (40–50 V) is shared between both the positive column and anode region.

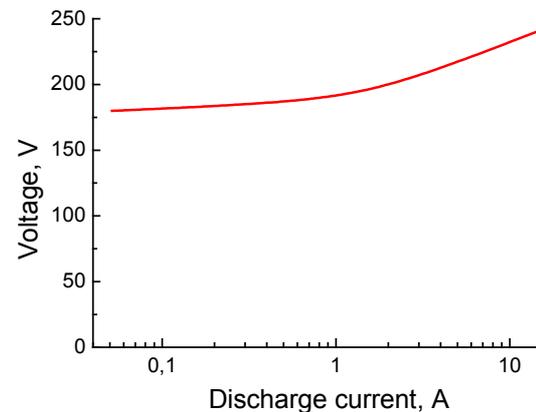

**Fig. 2.** *Voltage-current characteristic of the APGD in helium*

Schematic diagram of experimental apparatus for an absorption line profile registration is shown in fig. 3. The discharge image was focused on entrance slit of a double grating monochromator (MDD 500x2) of high resolution. A Gaussian instrumental profile was ~ 0.01 nm. To determine the concentrations of the excited helium atoms the absorption spectroscopy method was used [8]. The APGD was highlighted parallel to the cathode surface by a probing emission. A halogen incandescent lamp KGM-12-50 was used as probing light source for getting the absorption lines in visible spectral region. The original light source on the base of an arc with thermionic cathode [9] was used in the UV region. A probing emission was focused in plasma volume located on the discharge axes and than was collected on entrance slit of monochromator together with a discharge emission. To obtain the needed resolution in axial direction two slit-diaphragms were used. The electrical signal proportionate to the probing emission intensity was separated from a common photomultiplier signal due to both the modulation of the probing emission and the using of a selective amplifier. To make better a signal/noise ratio the synchronous detection was used as well.

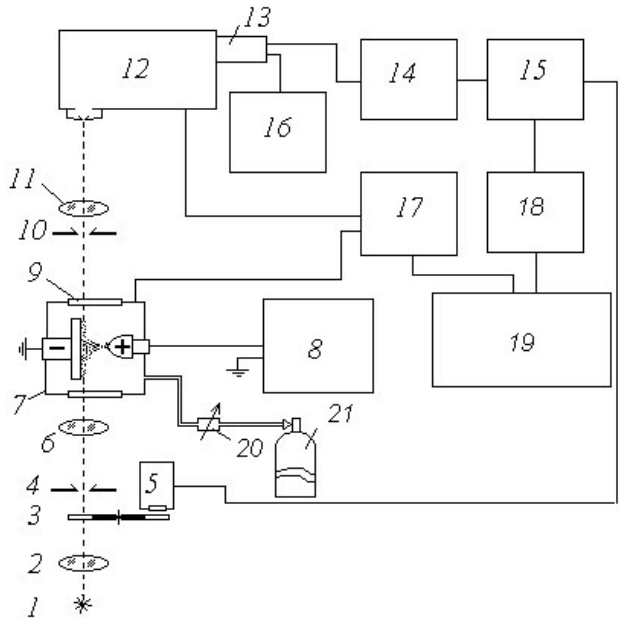

**Fig.3.** *Schematic diagram of the experimental apparatus.*
*1 – probing emission source, 2, 6, 11 – lens, 3 – chopper, 4, 10 – slit-diaphragms, 5 -- synchronous sensor, 7 discharge chamber, 8 – discharge power supply, 9 – windows, 12 – monochromator, 13 – photomultiplier, 14 – selective amplifier, 15 – synchronous detector, 16 – high-voltage power supply, 17 – control block, 18 – A/D converter, 19 – computer, 20 – flowmeter, 21 – gas-cylinder.*

The spectra of the probing emission before and after passing through plasma were registered in the experiments. Calculation of the excited helium atom concentrations in states $2^1S$, $2^1P$, $2^3S$ and $2^3P$ was fulfilled by numerical integration of the measured absorption line profiles at the wavelength 501.6 nm, 667.8 nm, 388.9 nm and 587.6 nm, correspondingly [8]. An axial concentration distribution was obtained due to a moving of discharge chamber by step motor in the discharge axes direction.

### III. EXPERIMENTAL RESULTS

Experiments were performed at discharge current 1 A. This value of discharge current was chosen due to a few causes. First, the sufficient cooling of cathode is provided in this case. And second, the diameter of negative glow consist of about 8 mm. That allows to obtain a needed spatial resolution in cathode region at side-on observation. Normal current density was about 2 A/cm$^2$.

In fig. 4a the dependence of the interelectrode voltage on nitrogen bulk concentration in helium flow is presented. Measurements were performed at interelectrode gap ~2.5 mm. As it is seen the interelectrode voltage is about 195 V when the helium flow does not contain a nitrogen admixture. This voltage value is pictured in fig. 4a by the dotted line. But the adding of small quantity of nitrogen in helium flow leads to voltage growth. So at nitrogen bulk concentration of 5% the voltage is equal to about 330 V. That exceeds on 135 V the interelectrode voltage observed in the APGD in helium. It is necessary to notice, current density almost does not change at nitrogen bulk concentrations less 5%, but it increases a few at higher ones.

The increase of interelectrode voltage takes place non-uniformly along the interelectrode gap.

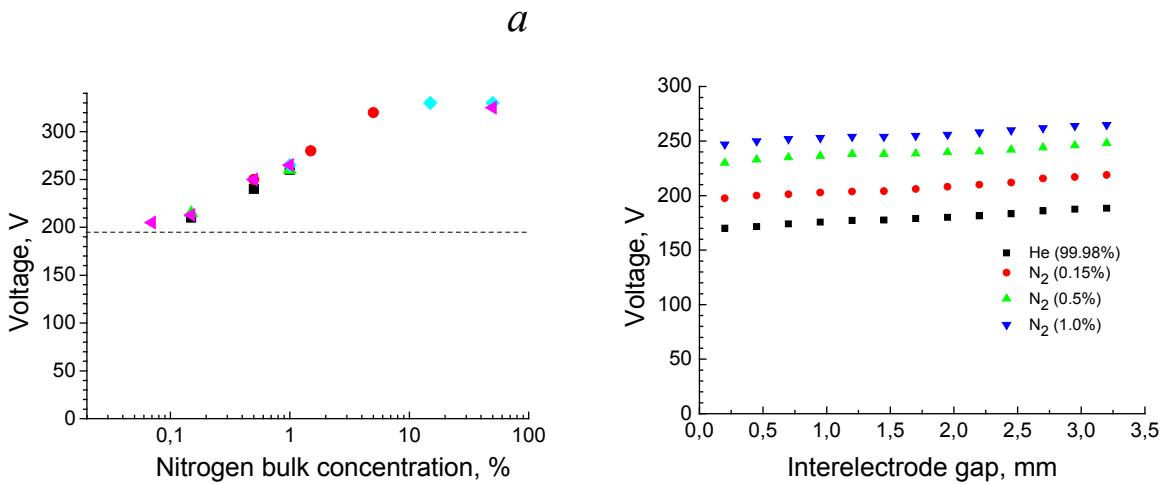

**Fig. 4.** *Dependences of interelectrode voltage on both (a) the nitrogen bulk concentration and (b) interelectrode gap at different nitrogen bulk concentrations.*

The voltage increase occurs in both cathode and anode region. This fact is confirmed by fig. 4b, where the voltage dependence on an interelectrode gap at different nitrogen bulk concentrations is presented. A change of gap does not lead to the visible changing in neither the cathode nor anode region. But in the case of the gap decrease the reducing of the positive column takes place. There is no positive column at interelectrode gap less than 0.5 mm. This confirms that the voltage increase is connected with the near-electrode regions, especially, with the cathode region. Probably, the changes in positive column at nitrogen adding are not considerable, because the obtained experimental data are located on parallel lines (Fig. 4b). As it seen in fig. 4b the cathode fall value in helium APGD is close to one in the lower pressure glow discharge [7]. At adding of nitrogen in helium the cathode fall increases. At nitrogen bulk concentration about 0.5% cathode fall is already exceeding significant the one in pure nitrogen glow discharge.

The significant increase of voltage drop takes place in cathode region including both the cathode fall region and negative glow. The thickness of this region in axial direction consists of 0.2-0.3 mm. The voltage drop increase in such thin layer should lead to considerable growth of volumetric heat generation and, as a consequence, a gas temperature increase. The gas temperature in the APGD was determined in experiments using the molecular bands in emission spectrum. The emission spectrum of the APGD plasma mainly consists of intensive lines of neutral helium atoms. However the more weak lines of the hydrogen, nitrogen, oxygen and other elements are observed as well (see fig 5a, for example). The $N_2^+(B^2\Sigma_u^+ - X^2\Sigma_g^+)$ electron-vibration bands (1,0) 358.2 nm,

(0,0) 391.4 nm, (0,1) 427.8 nm and (0,2) 470 nm of the first negative system of nitrogen are registered. The ($A^2\Sigma^+$-$X^2\Pi_i$) electron-vibration bands (0,0) 308.nm, (1,1) 314,3 of hydroxyl are observed in UV region. More intensive spectra are registered in negative glow.

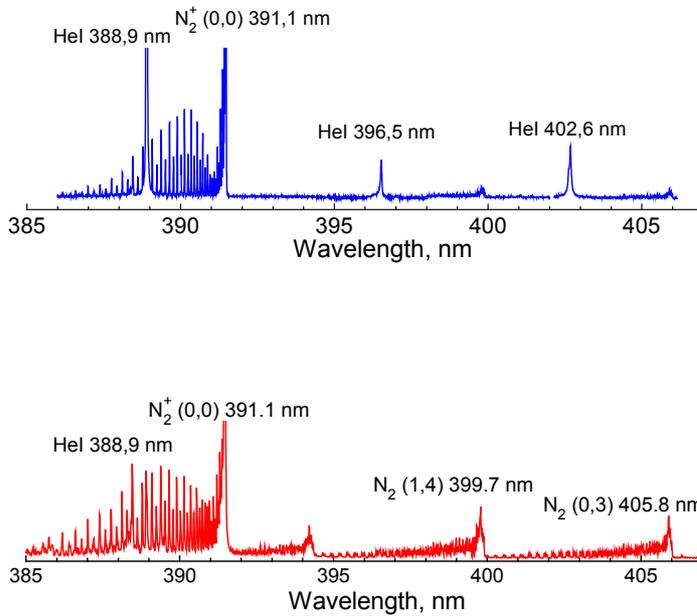

**Fig. 5.** *Segment of emission spectrum of the APGD plasma (a) in helium and (b) in helium-nitrogen mixture.*

The band (0,1)($\lambda$ = 427.81 nm) from the first negative system of molecular nitrogen ion $N_2^+(B^2\Sigma_u^+)$ served as a working band. The technique of determining the gas temperature $T_g$ is based on the measurements of the relative intensities of corresponding bands under the assumptions that the rotational level populations of molecules $N_2^+$ in the excited electronic state obey Boltzmann distribution and that the measured rotational temperature $T_{rot}$ corresponds to the gas temperature $T_g$ [10].

The gas temperature $T_g$ determined using the relative intensities of the $N_2^+$ band in negative glow of the APGD in helium is about 720 K. The adding of nitrogen into working gas helium leads to an increase of intensities of the electron systems of molecular nitrogen (Fig. 5b). This spectrum was registered in mixture containing 5% of nitrogen. If we compare these two spectra we will find the two other important differences. Firstly, it is obvious that the relative distribution of intensities of the $N_2^+$ band lines changed. The increase of intensities of molecular ion band lines with large number shows the increase of gas temperature. And really, the gas temperature in negative glow increased up to 860 K and 1055 K at values of nitrogen bulk concentrations 0.5% and 5% correspondingly.

A second interesting fact is that the intensity of the second positive system of nitrogen increases more quickly in comparison with one of the first negative system of nitrogen while concentration of nitrogen in mixture is increased. The bands of the second positive system of nitrogen are not practically observed in the helium APGD (Fig. 4a). At the same time the difference between the band intensities of both the first negative system and the second positive system of nitrogen becomes less in the APGD in helium-nitrogen mixture (Fig. 4b). It is necessary to notice that the intensities of nitrogen bands are increasing quickly (in one hundred times) at increase of nitrogen bulk concentration up to 0.2% (Fig. 5a). Therefore, a significant population growth of both the $N_2^+(B^2\Sigma^+)$ and the $N_2(C^3\Pi_u)$ levels takes place in the APGD in helium-nitrogen mixture. At higher bulk concentration the intensities of nitrogen bands are changed slightly.

*a*    *b*

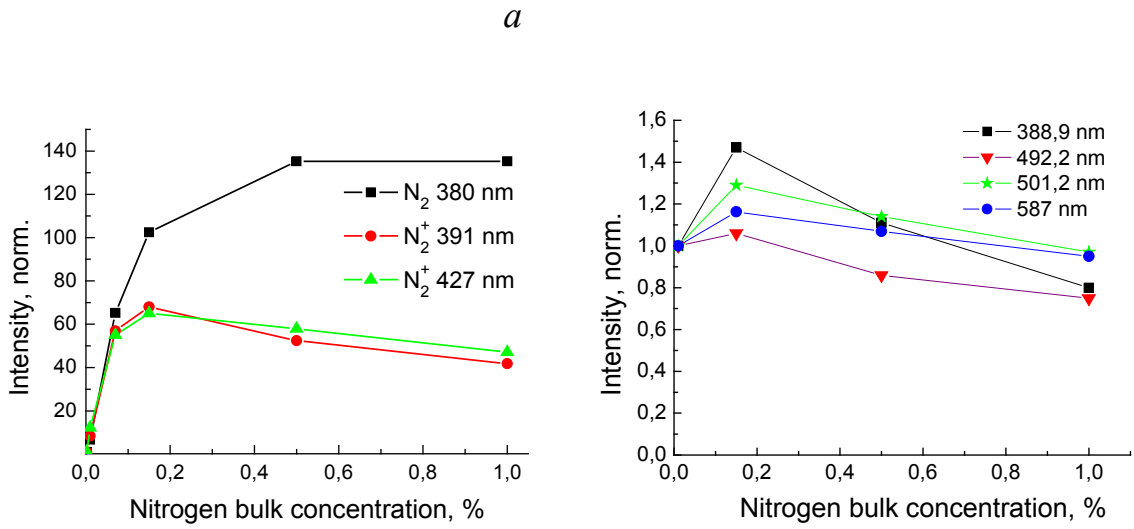

The intensities of helium atom lines in mixture are changed in limits of 20% from its intensity in the helium APGD (Fig. 5b). At the same time the adding of nitrogen in helium influences on populations of the low-excited helium atom levels (n = 2). The transmittance line profiles at the wavelength 501.6 nm, 667.8 nm, 388.9 nm и 587.6 nm are presented in fig. 6. These profiles were registered in negative glow. It is seen at the adding of the 0.5% nitrogen in helium a significant decrease of an absorption is observed at all lines.

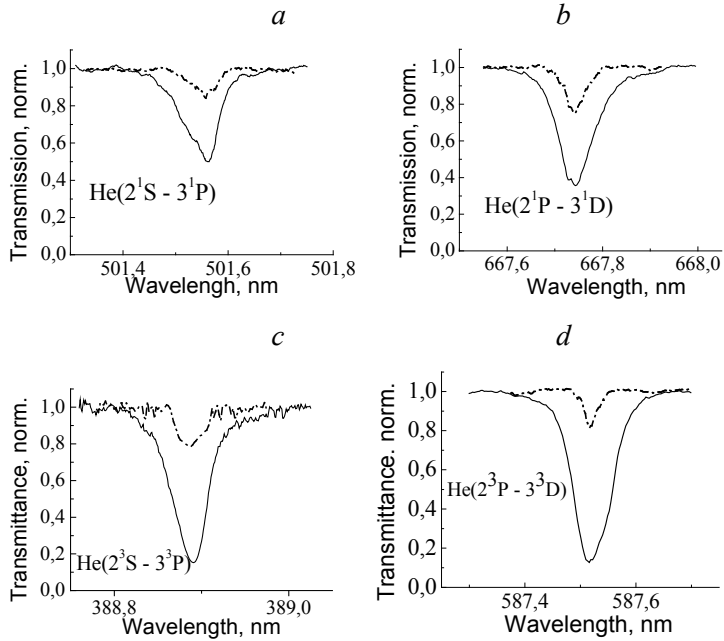

**Fig. 6.** *The transmission line profiles at the wavelength 501.6 nm (a), 667.8 nm (b), 388.9 nm (c) и 587.6 nm (d), solid curve – discharge in helium, dashed one in mixture.*

Using the experimental transmission profile (Fig. 6) the concentration of atoms $N_n$ in the corresponding state can be calculated as follows [8]

$$N_n \approx c/(0.026 f_{nm} l\lambda_0^2) \int \ln(I_0/I_\lambda)d\lambda, \qquad (1)$$

where $\lambda_0$ – wavelength of corresponding line, $l$ – thickness of absorption layer which was supposed to be homogeneous, $c$ – light velocity, $f_{mn}$ – oscillator force.

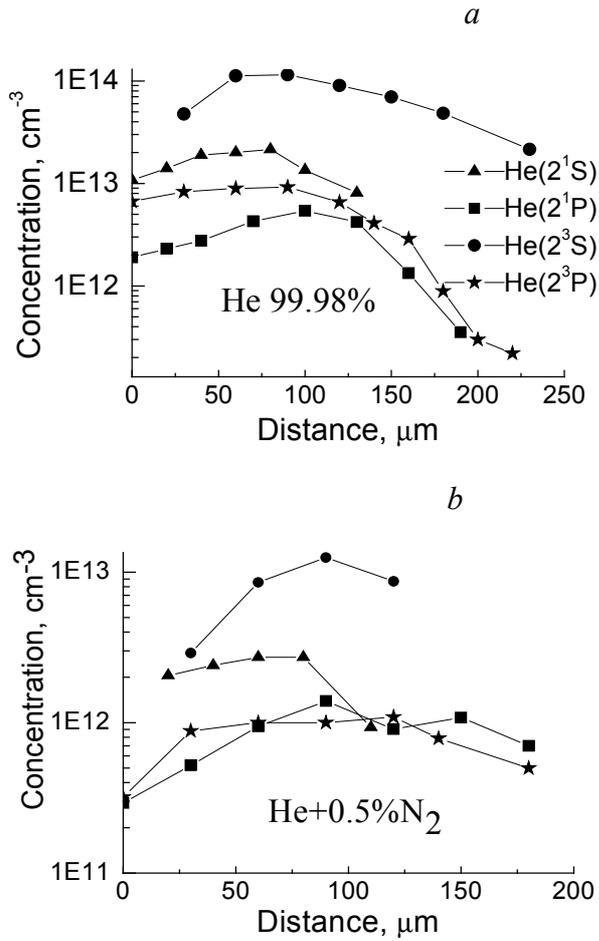

**Fig. 7.** *Axial distributions of the exited helium atom concentrations in helium discharge (a) and in helium mixture discharge (b)*

The transmittance profiles were registered at different distances from cathode. Using these experimental profiles and formula (1) the corresponding axial distributions of concentration of low-excited helium atoms were obtained. They are presented in fig. 7. It can be seen that the concentration of the excited helium atoms in helium-nitrogen mixture (99.5%He + 0.5%N$_2$) is less one order of magnitude than in case of pure helium. The axial concentration distributions are analogous for discharges in helium and helium-nitrogen mixture. Maximal concentration magnitude takes place in cathode region at distance about 0.1 mm from cathode surface.

Using an experimental data for both spectral intensities and transmittances (Fig. 6) it is possible to receive the relative populations of radiative levels of helium atoms. A solution of the radiative transfer equation for homogeneous layer at absence of incident external emission it can write as follows

$$I_\lambda(x=l) = \frac{-l(1-T_\lambda)}{\ln T_\lambda} J_\lambda .$$

Here $I_\lambda(x=l)$ - intensity of emission emanating out of layer with thickness $l$, $J_\lambda$ - plasma emission coefficient, $T_\lambda$ - layer transmittance. An integral emission coefficient $J = \int J_\lambda d\lambda$ is connected with population of upper radiative level $N_n$ by expression

$$J = \frac{g_m}{g_n} \frac{2\pi\, e^2 h v^3}{mc^3} f_{mn} N_n ,$$

where $g_m$ and $g_n$ - statistical weight of lower and upper levels. Finally for concentration of helium atoms on upper radiative level $N_n$ it is possible to receive using a follow expression:

$$N_n = -\left(\frac{g_n}{g_m}\right)\frac{\lambda_0^3 m}{2\pi e^2 h}\frac{1}{l}\int\frac{\ln T_\lambda}{(1-T_\lambda)} I_\lambda(x=l)\, d\lambda.$$

In fig. 8 the relative concentration dependences of helium atoms in the excited states $2^1S$ and $3^1P$ determined using both the absorption and emission profiles of the HeI 501.6 nm line in negative glow on a nitrogen bulk concentration are shown. Normalization was fulfilled on concentration of these kinds of atoms in the helium APGD. It is seen that at adding of nitrogen in helium the He ($3^1P$) atom concentration decreases slightly. At the same time the He ($3^1S$) atom concentration decreases drastically (one order of magnitude) already at adding of 0.5% nitrogen. This implyies that the main reason of a decrease of lower-excited He$^*$(n=2) atom concentration is a quenching these excited atoms at their collisions with nitrogen molecules.

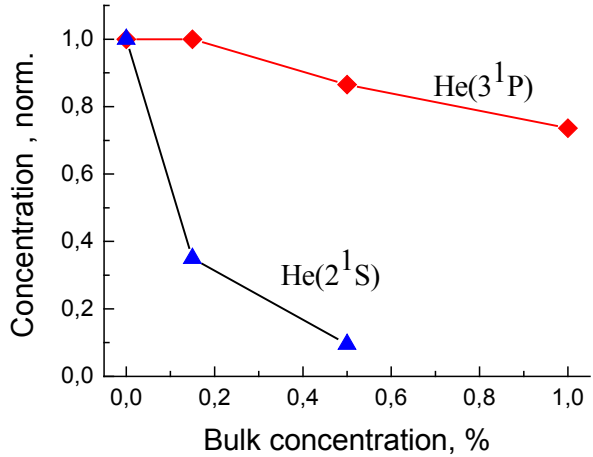

**Fig. 8.** *Relative concentration of both the $3^1P$ and $2^1S$ helium atoms against nitrogen bulk concentration.*

## IV. DISCUSSION

Let us make the analysis of the main reactions in the mentioned both APGDs. The low-exited levels of helium atoms play an important role in stepped ionization processes. The role of this ionization mechanism in the APGD is more significant in comparison with one in the lower pressure glow discharge. In the electric fields $E/N_0 \leq 10^{-15}$ V·cm$^2$ and at the ionization degree more $10^{-5}$ a stepped ionization in the APGD plasma is more important than a direct ionization by electron collision even at free exit of light emission [11]. Naturally, a light reabsorption increases the stepped ionization efficiency. The charged particles are produced in follow reactions with participation of excited helium atoms He$^*$ [12]:

$$\text{He}^* + \text{He}^* \rightarrow \text{He}^+ + \text{He} + e, \qquad (2)$$

$$\text{He}^*(n=3) + \text{He} \rightarrow \text{He}_2^+ + e. \qquad (3)$$

The scheme of reactions in gas mixture discharge should be added by the following reactions which describe an interaction of helium atoms and molecules with the nitrogen molecules [12-14]

$$He^* + N_2 \rightarrow He + N_2^+ + e, \quad (4)$$

$$He_2(^3\Sigma_u^+) + N_2 \rightarrow He + He + N_2^+ + e, \quad (5)$$

$$He^+ + N_2 \rightarrow He + N + N^+ + 0,3 \; eV, \quad (6)$$

$$He_2^+ + N_2 \rightarrow He + He + N_2^+. \quad (7)$$

Reaction (4) is in charge of a decrease of the excited helium atom concentrations at a nitrogen adding, because a decrease of the He$^*$ atom concentration due to the temperature growth is neglible. Since a significant part (a few tens of percents) of generated molecular ions of nitrogen is in the excited B$^2\Sigma^+$ state, the quenching of excited helium atoms is accompanied by emission in the first negative system of nitrogen. The generation of molecular nitrogen ion is a result of a quenching process of the He$_2$(2$^3\Sigma_u^+$) molecule at collision of metastable helium atom with nitrogen molecule (reaction 5). Reactions (6), (7) describe the processes of recharge at collision of both the atomic and molecular helium with nitrogen molecules.

Thus, the adding of nitrogen into helium leads to the reduction of concentration of both the low-excited helium atoms and ions. As result the nitrogen becomes in charge of the maintenance of the APGD in helium-nitrogen mixture even in presence of small nitrogen admixture.

The work was supported by BRFBR (grant T04-181).